\documentclass[%
 reprint,
superscriptaddress,
 amsmath,amssymb,
 aps,
]{revtex4-2}
\usepackage{xcolor, soul}
\sethlcolor{yellow}
\usepackage{graphicx}
\usepackage{dcolumn}
\usepackage{bm}
\usepackage{lipsum}

\usepackage{subfigure}

\begin{document}


\title{Enhancement of spin to charge conversion efficiency at the topological surface state by inserting normal metal spacer layer in the topological insulator based heterostructure}

\author{Sayani Pal}
\affiliation{Department of Physical Sciences, Indian Institute of Science Education and Research Kolkata, Mohanpur 741246, West Bengal, India}
\author{Anuvab Nandi}
\affiliation{Department of Physical Sciences, Indian Institute of Science Education and Research Kolkata, Mohanpur 741246, West Bengal, India}
\author{Shambhu G. Nath}
\affiliation{Department of Physical Sciences, Indian Institute of Science Education and Research Kolkata, Mohanpur 741246, West Bengal, India}
\author{Pratap Kumar Pal}
\affiliation{Department of Condensed Matter and Materials Physics, S. N. Bose National Centre for Basic Sciences, Salt Lake, Kolkata 700106, India}
\author{Kanav Sharma}
\affiliation{Department of Physical Sciences, Indian Institute of Science Education and Research Kolkata, Mohanpur 741246, West Bengal, India}
\author{Subhadip Manna}
\affiliation{Department of Physical Sciences, Indian Institute of Science Education and Research Kolkata, Mohanpur 741246, West Bengal, India}
\author{Anjan Barman}
\affiliation{Department of Condensed Matter and Materials Physics, S. N. Bose National Centre for Basic Sciences, Salt Lake, Kolkata 700106, India}
\author{Chiranjib Mitra}
\email{Corresponding author: chiranjib@iiserkol.ac.in}
\affiliation{Department of Physical Sciences, Indian Institute of Science Education and Research Kolkata, Mohanpur 741246, West Bengal, India}

\date{\today}

\begin{abstract}
We report efficient spin to charge conversion (SCC) in the topological insulator (TI) based heterostructure ($BiSbTe_{1.5}Se_{1.5}/Cu/Ni_{80}Fe_{20}$) by using spin-pumping technique where $BiSbTe_{1.5}Se_{1.5}$ is the TI and $Ni_{80}Fe_{20}$ is the ferromagnetic layer. The SCC, characterized by inverse Edelstein effect length ($\lambda_{IEE}$) in the TI material gets altered with an intervening Copper (Cu) layer and it depends on the interlayer thickness. The introduction of Cu layer at the interface of TI and ferromagnetic metal (FM) provides a new degree of freedom for tuning the SCC efficiency of the topological surface states. The significant enhancement of the measured spin-pumping voltage and the linewidth of ferromagnetic resonance (FMR) absorption spectra due to the insertion of Cu layer at the interface indicates a reduction in spin memory loss at the interface that resulted from the presence of exchange coupling between the surface states of TI and the local moments of ferromagnetic metal. The temperature dependence (from 8K to 300K) of the evaluated $\lambda_{IEE}$ data for all the trilayer systems, TI/Cu/FM with different Cu thickness confirms the effect of exchange coupling between the TI and FM layer on the spin to charge conversion efficiency of the topological surface state.

\end{abstract}

\pacs{Valid PACS appear here}
\maketitle

\begin{figure}
\centering
\includegraphics[scale=0.4]{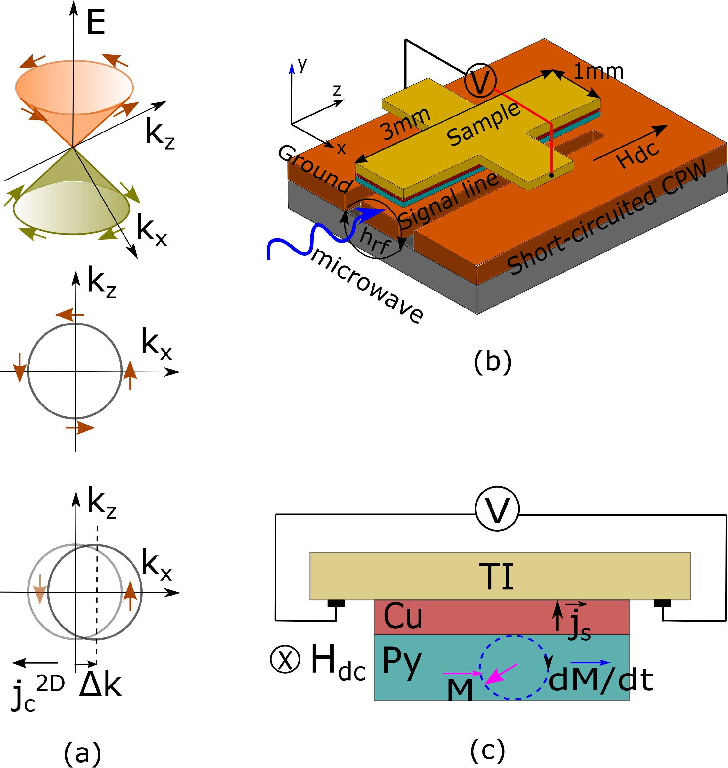}
\caption{Schematic illustration of the (a) concept of spin-to-charge conversion  at TSS. Top panel: The energy dispersion relation of TSS with helical spin configurations of opposite chirality above and below the Dirac cone. Middle panel: spin configuration of the Fermi surface of TSS. Bottom panel: depicts the IEE in the device. The injected spin density from FM layer, $j_s^{3D}$ with a specific polarity (here dark shaded orange arrow), accumulates at the TI surface state and creates an imbalance of spin population and generates a charge current, $j_c^{2D}$ on the plane of surface state of TI; (b) experimental set-up for spin-pumping measurements on $BiSbTe_{1.5}Se_{1.5}/Cu (t)/Ni_{80}Fe_{20}$ heterostructure and (c) cross-section of the sample at the FMR resonance condition when magnetization, $M$ precesses around the external dc magnetic field ($H_{dc}$) and due to the precession of magnetization ($dM/dt$), a spin current gets injected from the FM layer into the TI layer with spin-current density, $j_s$ and spin polarization along the direction of the $H_{dc}$ field.}
\label{Fig1}
\end{figure}

Materials with efficient spin-charge interconversion (SCI) in the realm of Spintronics is gathering increasing attention due to their capacity to facilitate the development of energy-efficient spin-based devices. Enhanced SCI can be achieved by harnessing the spin orbit coupling (SOC) induced effect in different systems. Spin Hall SOC in heavy metals, Rashba SOC in two-dimensional electron gas (2DEG) at certain surfaces and interfaces and spin-momentum locked surface states in topological insulators (TI) yield substantial SCI capabilities. Recent experiments have observed remarkable spin-charge interconversion (SCI) in heterostructures based on TI \cite{Optical,Pumping,FI}. One versatile method of SCI that stands out is the spin pumping experiment \cite{Suchetana, SPump, geff}. This process generates a spin current through the precession of magnetization in a ferromagnetic layer. The resulting spin current polarization aligns with the precession axis and induces a spin imbalance on the TI surface as the accumulated spin polarization diffuses into the TI. This imbalance, owing to the unique spin texture of TI state known as spin-momentum locking, leads to a spin accumulation and the creation of an electric field along the transverse direction via inverse Edelstein effect (IEE) as demonstrated in the Fig.\ref{Fig1}a. The parameter known as IEE length, $\lambda_{IEE}$ involves the conversion of spin accumulation at the TI/FM interface into a transverse electric field in the TI and it quantifies the strength of spin to charge conversion (SCC) within the system. $\lambda_{IEE}$ for the TI surface state can be expressed as $\lambda_{IEE} = j_c^{2D}/j_s^{3D} = v_F \tau$ where $j_s^{3D}$ is the injected spin current density from FM and $j_c^{2D}$ is the generated charge current density in TI respectively. $v_F$ is the Fermi velocity of the Dirac fermions and $\tau$ is the relaxation time of the non-equilibrium spin density distribution at the TI surface state \cite{Fert1}. IEE at the topological surface state (TSS) is found to be more efficient in SCI than the conventional inverse spin Hall effect in heavy metals\cite{HM1,HM2,HM3}. However, the TI/FM interface may have significant effect on the SCI efficiency of the device in the following ways: {\romannumeral 1)} spin density from Rashba-split 2DEGs at the TI/FM interface \cite{BB2}, originated from band bending \cite{BB1} can negatively affect the SCI of the device; {\romannumeral 2)} presence of exchange coupling between TI surface state and FM can introduce spin memory loss at the interface \cite{S2} and, {\romannumeral 3)} material intermixing \cite{L2} at the interface of TI and FM can lead the formation of hybridization state and magnetic dead layer \cite{MD1}. These deviations from an ideal TI/FM interface can play a critical role in suppressing the SCI efficiency. Thus for future TI-based spintronic devices we need a complete understanding of the interface effect on the efficiency of the devices.\\
In this paper, we demonstrate the spin-electricity conversion at the TI surface state and at the interface of TI with ferromagnetic and non-magnetic metals. To achieve this, we examined a specific heterostructure, $BiSbTe_{1.5}Se_{1.5}/Cu/Ni_{80}Fe_{20}$, by performing the ferromagnetic resonance (FMR) and spin-pumping measurements. We chose $BiSbTe_{1.5}Se_{1.5}$ as the 3D TI material where the surface state conduction dominates even at room temperature, thanks to the negligible bulk state contribution in conduction\cite{BSTS1,BSTS2}. We took Cu as the spacer layer between $BiSbTe_{1.5}Se_{1.5}$ (BSTS) and $Ni_{80}Fe_{20}$(Py) because Cu has long spin diffusion length and the Cu/Py interface has higher spin scattering asymmetry. The Cu/Py interface exhibit strong spin filtering due to weak majority-spin scattering and strong minority spin scattering at the interface \cite{Cu_Py_interface1}.\\
We have prepared the heterostructure, $BiSbTe_{1.5}Se_{1.5}/Cu/Ni_{80}Fe_{20}$, with varying thicknesses of the copper layer (0.5 nm, 1.5 nm, and 3 nm). The thickness of the BSTS layer and the Py layer remained constant at 37 nm and 20 nm, respectively. Dimensions of the sample has been described in detail in the schematic diagram shown in Fig.\ref{Fig1}b. For clarity, we designated the different thicknesses of Cu as follows: Cu1 for 1.5 nm Cu and Cu2 for 3 nm Cu.                                                                                                                                                                                                                                                                                                                                                                                                                                                                                                                                                                                                                                                                                                                                                                                                                                                                                                                                                                                                                                                                                                                                                                                                                                                                                                                                                                                                                                                                                                                                                                 
The schematic diagram shown in Fig.\ref{Fig1}b illustrates the arrangement for FMR and spin pumping measurements. The sample has been placed on top of a coplanar waveguide having signal line width, 900 $\mu m$ and characteristic impedance of 50$\Omega$ \cite{Pal1}. An external magnetic field, ($H_{dc}$) in the plane of the sample and a microwave field ($h_{rf}$) of frequency 5$GHz$, perpendicular to $H_{dc}$ has been applied. The magnetization precession inside the FM layer at ferromagnetic resonance condition pumps spin angular momentum into the TI layer and it results in a spin accumulation at the TI surface state which in-turn develops into a transverse voltage across the TI in the microVolt range. The spins remain confined within the interface plane due to the two-dimensional nature of the topological surface state (TSS) and move perpendicular to the direction of spin polarization.

\begin{figure}
\centering
{\includegraphics[scale=0.172]{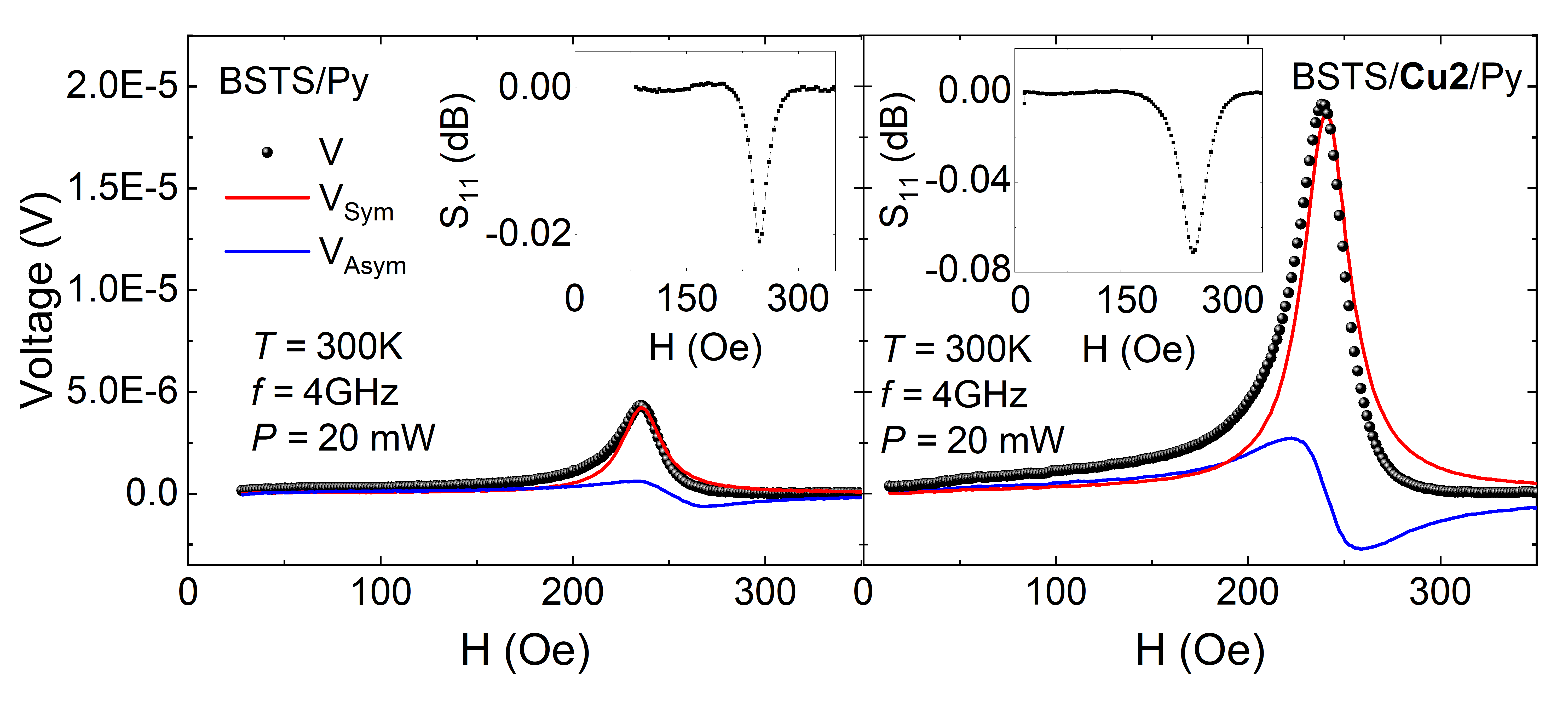}}
\caption{Spin-pumping voltage (V) measured for the samples $BSTS/Py$ and $BSTS/Cu(3nm)/Py$ at room temperature where $V_{Sym}$ and $V_{Asym}$ represents the symmetric and anti-symmetric component of the measured voltage at microwave power (P) of 20mW. The figures in the inset presents the typical FMR spectra at room temperature at resonance frequency of 4GHz.}
\label{Fig2}
\end{figure}

In Fig.\ref{Fig2}, the transverse voltage signal (V) as a function of external magnetic field (H) and the typical FMR resonance spectra (please refer to the inset figures) has been shown for BSTS/Py and BSTS/Cu(3nm)/Py respectively. The voltage measured between the two electrodes can be described by two components: the spin-pumping voltage signal that has the form of a symmetric Lorentz function, $V_{Sym}$ and the voltage contribution from the possible anomalous Hall effect (AHE) and anisotropic magnetoresistance (AMR) of the magnetic layer (Py), $V_{Asym}$ which has the form of antisymmetric Lorentz function\cite{Vasym1,Vasym2}. By fitting the measured voltage to the form $V= V_{Sym}(\Delta H^2/(\Delta H^2+(H-H_{res})^2))+V_{Asym}((\Delta H(H-H_{res}))/(\Delta H^2+(H-H_{res})^2))$, the $V_{Sym}$ component has been extracted for further evaluation of parameters useful for studying spin to charge conversion (SCC) at the TI surface state (TSS).

\begin{figure}
\centering
\subfigure[]{\includegraphics[scale=0.172]{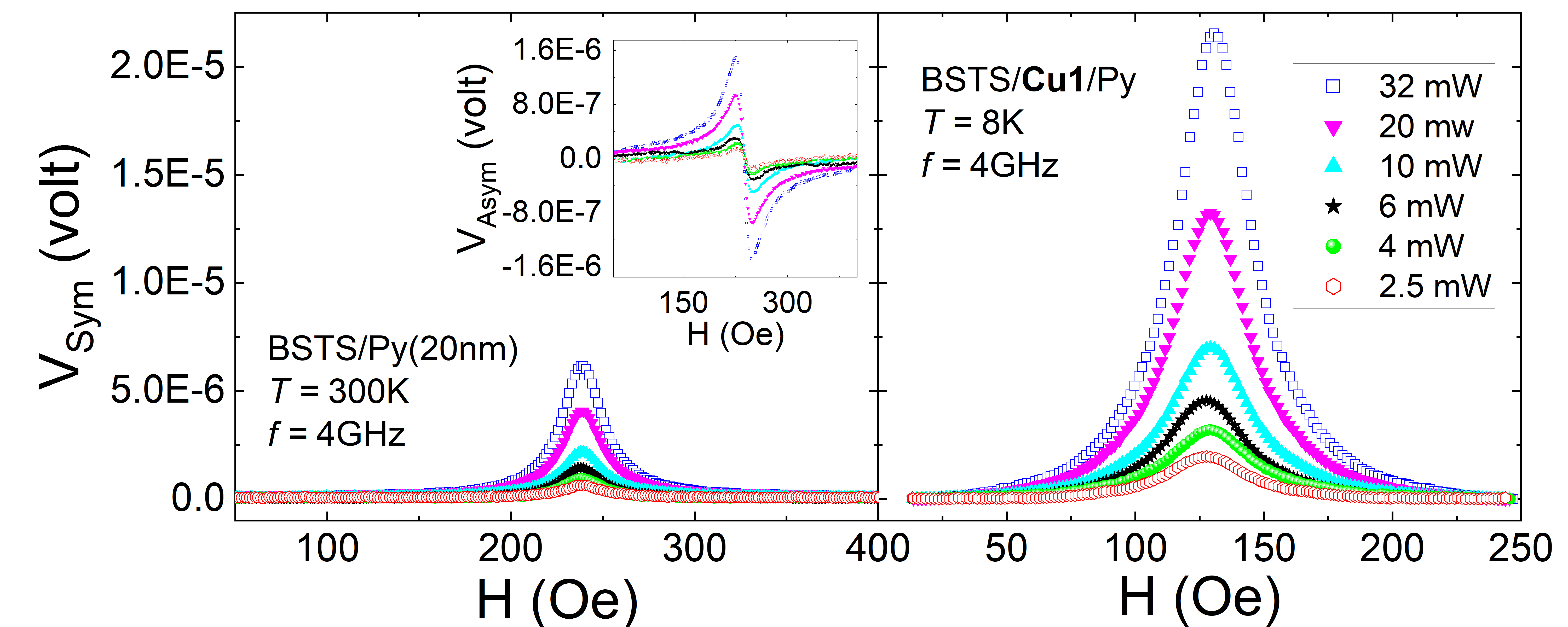}}
\subfigure[]{\includegraphics[scale=0.15]{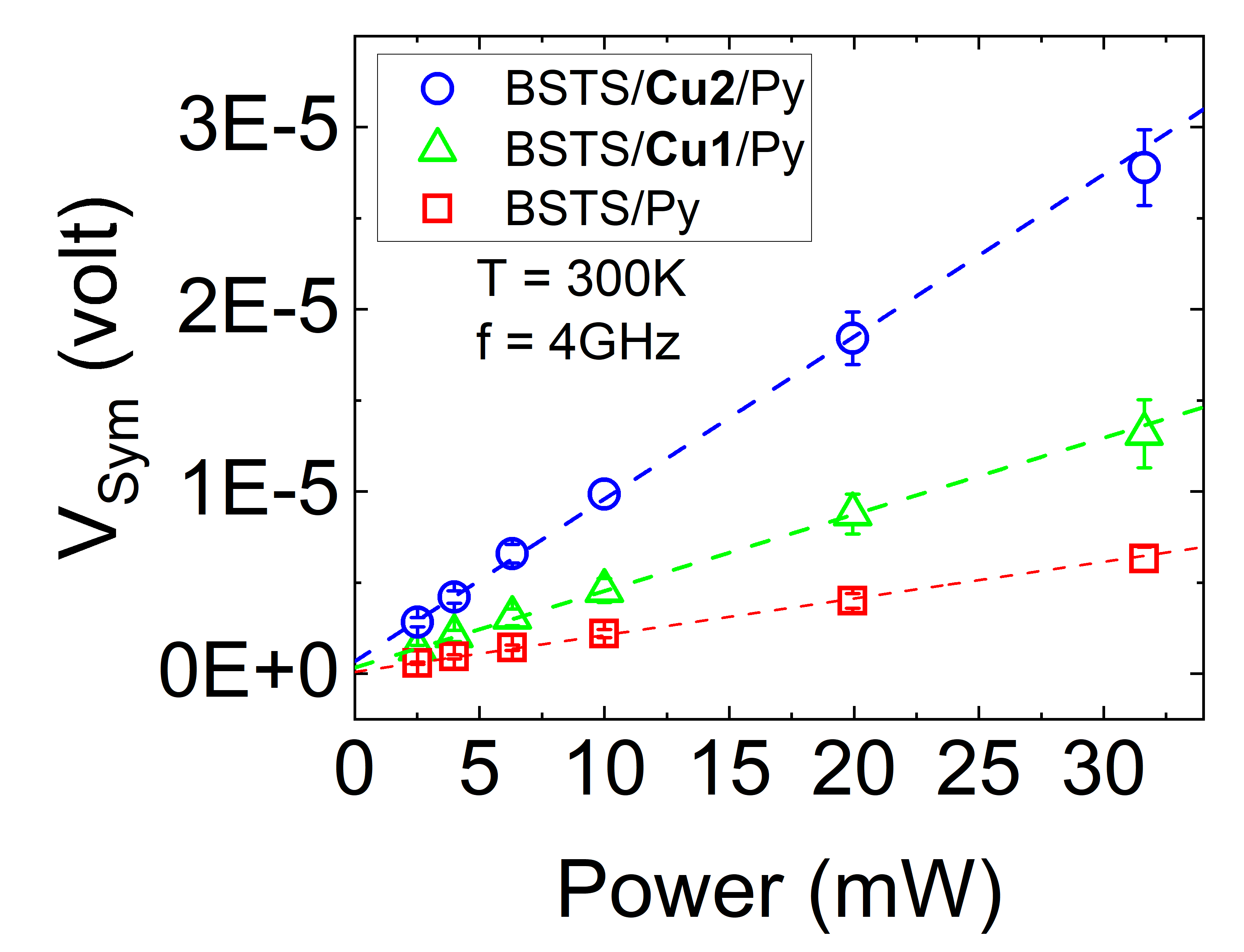}}
\subfigure[]{\includegraphics[scale=0.15]{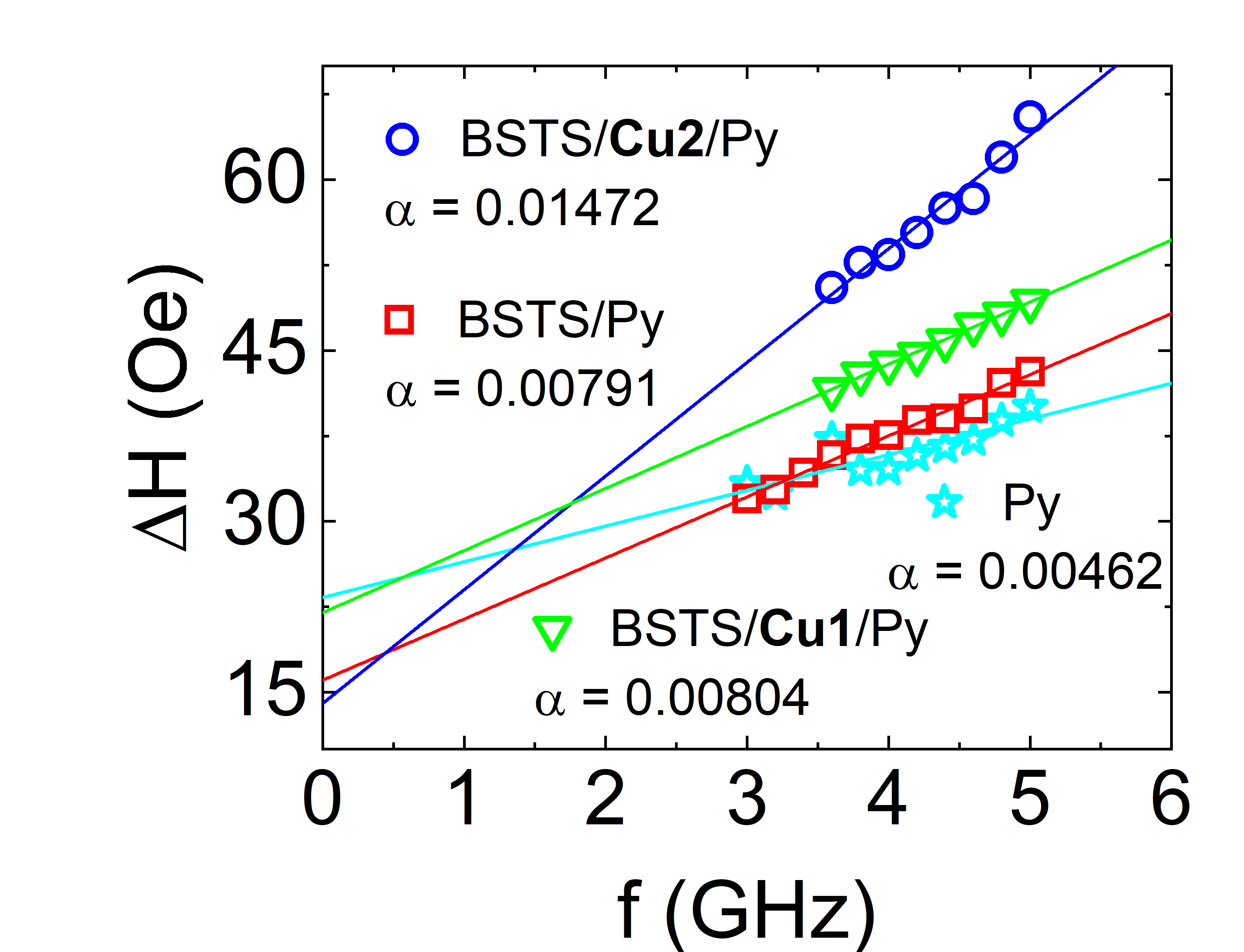}}
\caption{(a) Measured spin-pumping voltage ($V_{Sym}$) of the samples, $BSTS/Py$ and $BSTS/Cu(1.5 nm)/Py$ with varying microwave power at room temperature and at 8K respectively; The inset in the left figure shows $V_{Asym}$ with varying power; (b) $V_{Sym}$ varies linearly with the applied microwave power for all the heterostructure; (c) Linear fitting of FMR linewidth ($\Delta H$) vs frequency (f) and the evaluated damping coefficient ($\alpha$) values is shown for all the samples.}
\label{Fig3}
\end{figure}

\begin{figure}
\centering
\subfigure[]{\includegraphics[scale=0.15]{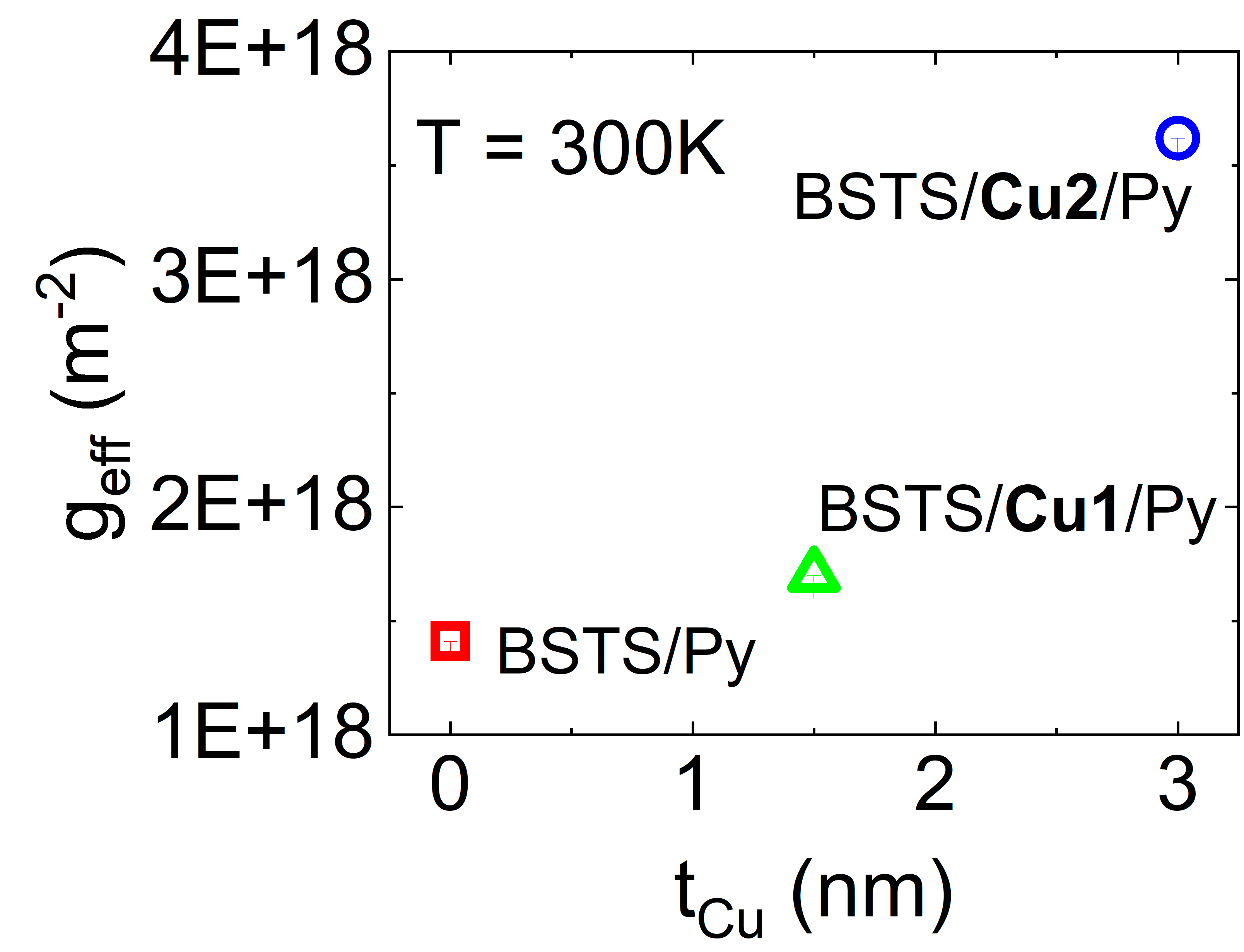}}
\subfigure[]{\includegraphics[scale=0.15]{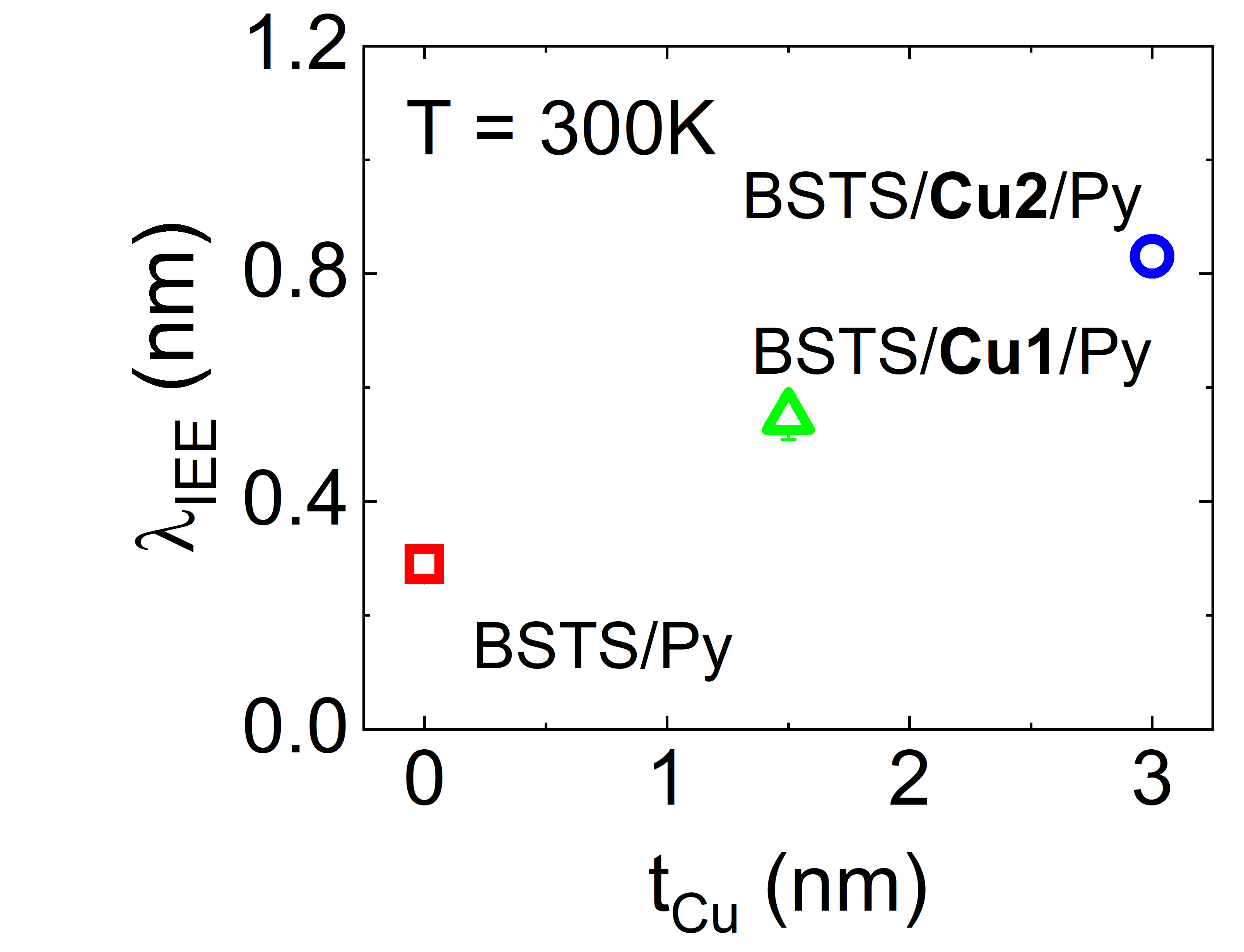}}
\caption{Obtained (a)spin-mixing conductance ($g_{eff}$) and (b) spin to charge conversion length ($\lambda_{IEE}$) with varied Cu thickness at room temperature.}
\label{Fig4}
\end{figure}

\begin{figure}
\begin{center}
\subfigure[]{\includegraphics[scale=0.23]{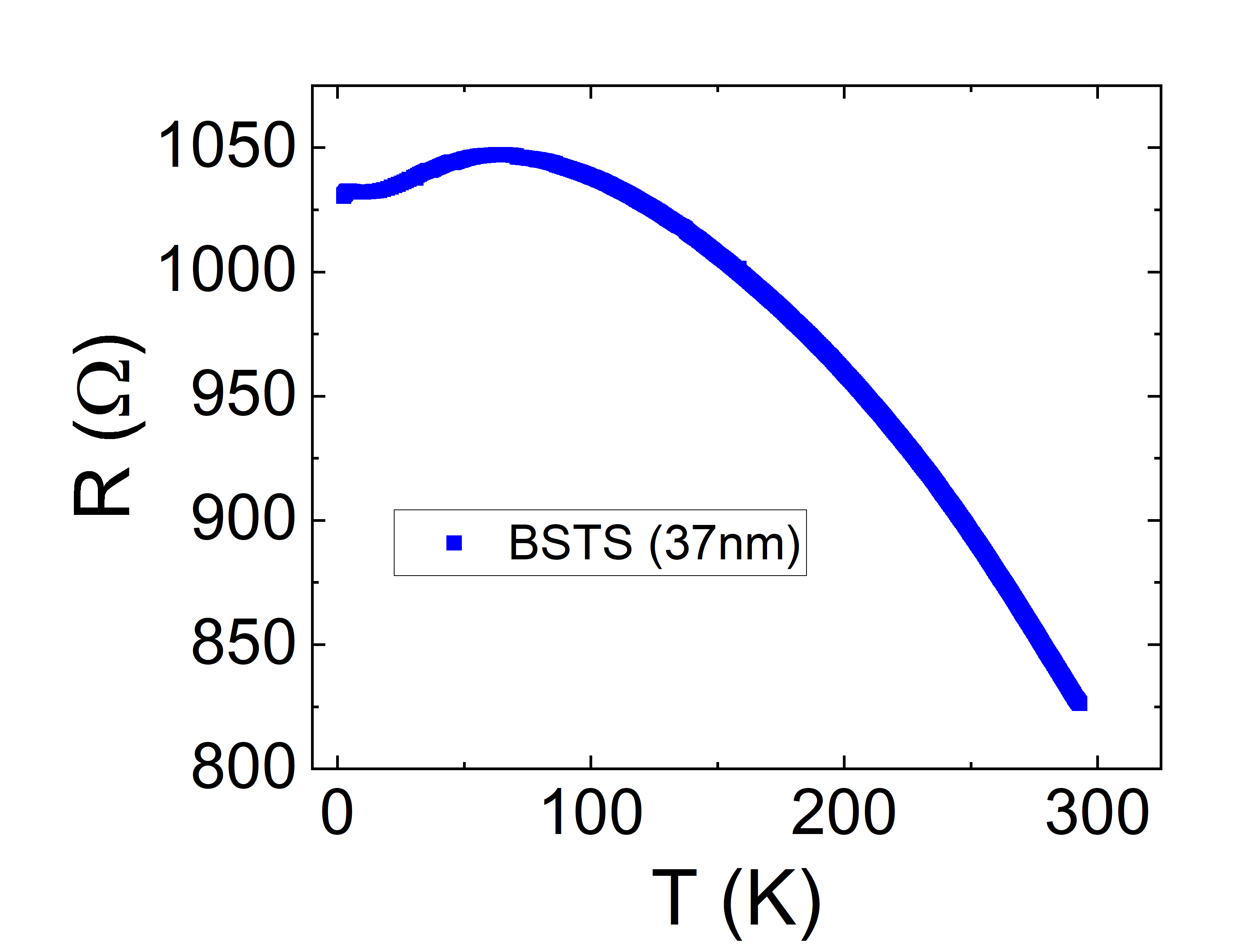}}
\subfigure[]{\includegraphics[scale=0.23]{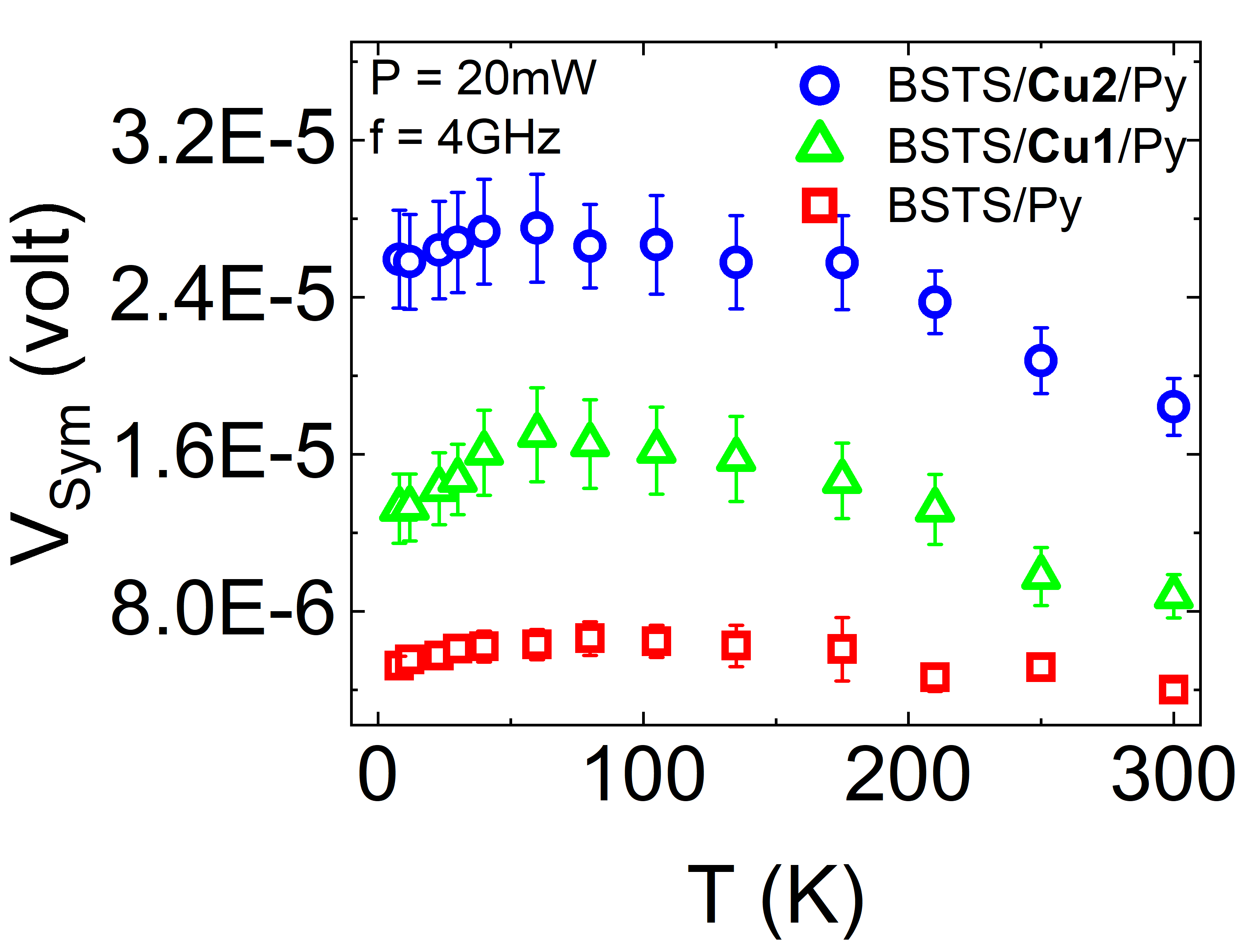}}
\subfigure[]{\includegraphics[scale=0.23]{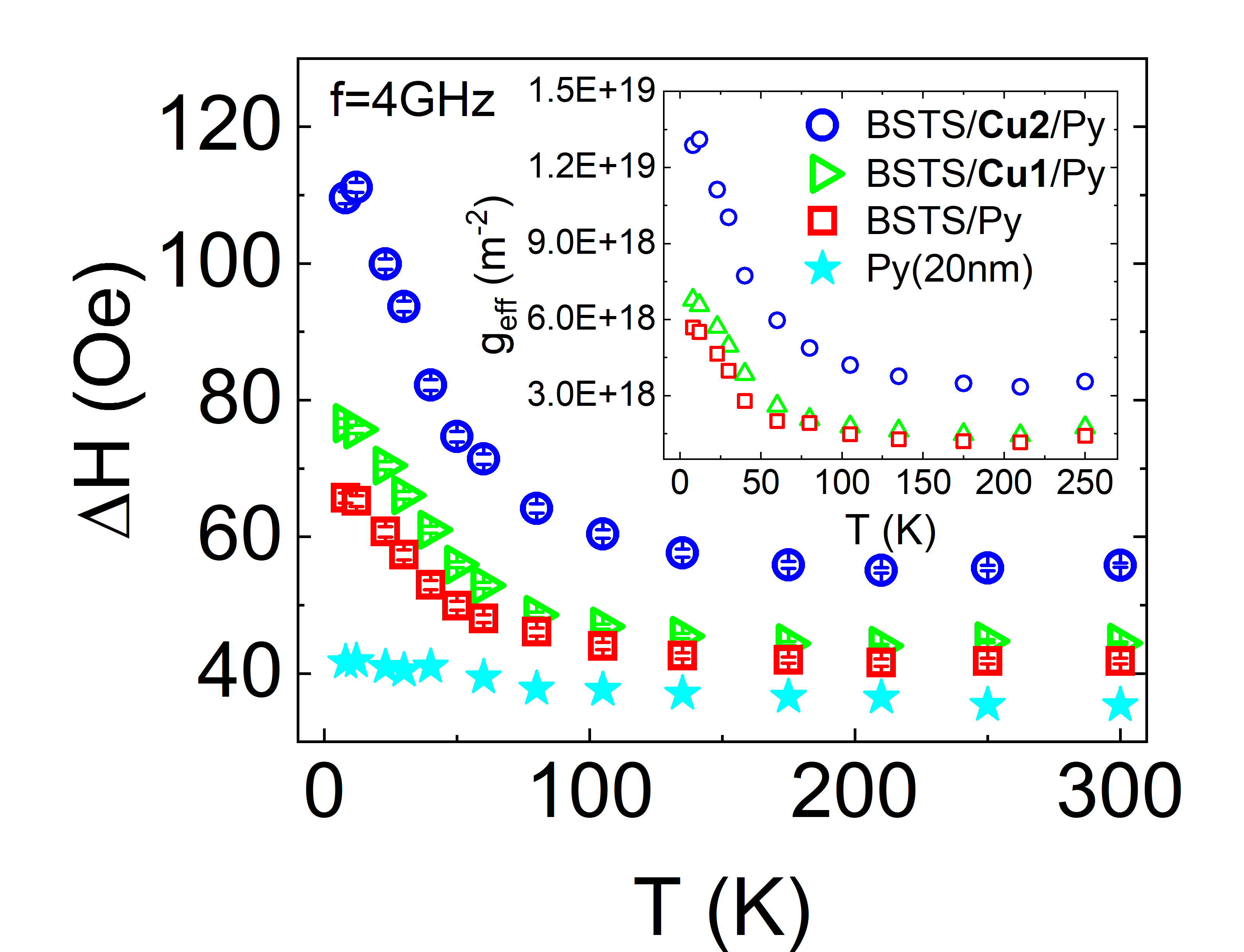}}
\subfigure[]{\includegraphics[scale=0.3]{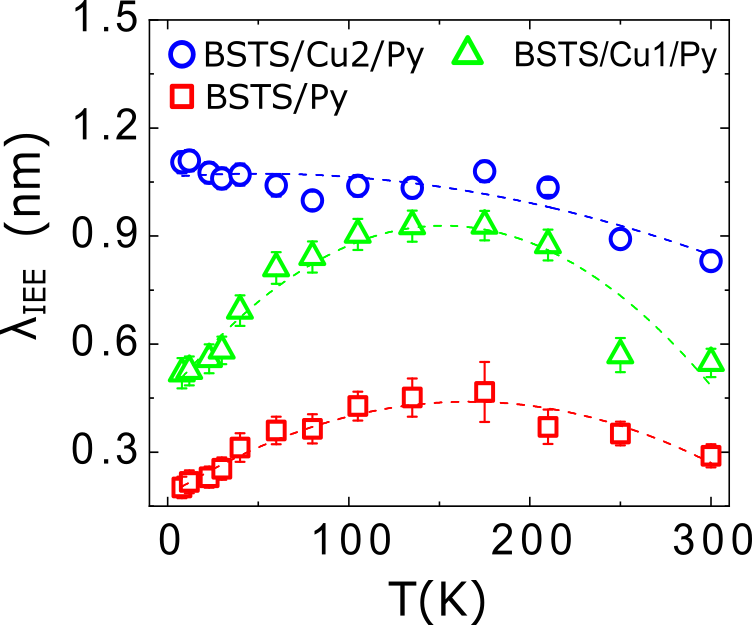}}
\end{center}
\caption{Temperature dependence of (a) Resistance ($R$) for 37 nm $BiSbTe_{1.5}Se_{1.5}$ thin film, (b) FMR spectra linewidth, $\Delta H$ (and spin-mixing conductance, $g_{eff}$ in the inset), (c) Symmetric component of the measured voltage, $V_{Sym}$  and (d) spin-to-charge conversion length ($\lambda_{IEE}$) for all the samples.}
\label{Fig5}
\end{figure}

In Fig.\ref{Fig3}a, the microwave power response of $V_{Sym}$ for $BSTS/Py$ and $BSTS/Cu2/Py$ has been shown at the excitation frequency of 4 GHz at room temperature and at low temperature respectively. The symmetric component of the measured voltage varies linearly with the microwave power for all the samples as can be seen from Fig.\ref{Fig3}b. The linearity suggests that the measured voltage is induced by the inverse Edelstein effect (IEE) as a result of spin pumping from the FM layer into the TI surface state\cite{L1,L2,L3}. We can observe from Fig.\ref{Fig3}b that the introduction of the Cu layer at the interface between the BSTS and Py layer leads to a general rise in the value of $V_{Sym}$. Additionally, in Fig.\ref{Fig3}c, we can see that the trilayer device exhibits an enhancement in the damping coefficient ($\alpha$) value compared to the single layer Py film and $BSTS/Py$ bilayer. It depicts the fact that the insertion of Cu layer definitely enhances the spin pumping efficiency in the device. We have calculated the effective spin mixing conductance, $g_{eff}^{\uparrow \downarrow}$ which is a material parameter that governs the transport of spin current into the adjacent NM layer from the FM layer and is proportional to the torque acting on the FM in the presence of spin accumulation in the NM layer \cite{bra1, bra2}. In Fig.\ref{Fig4}a, the variation of effective spin-mixing conductance ($g_{eff}$) with Cu thickness shows an increase in the $g_{eff}$ value for BSTS/Cu/Py samples compared to BSTS/Py bilayer. $g_{eff}$ was evaluated from the relation $g_{eff} = (2\sqrt{3}\pi M_{eff}\gamma d_{FM}(\Delta H_{bi}-\Delta H_{fm}))/(g\mu_b \omega)$ \cite{geff}. Given that $g_{eff}$ takes into account the spin back flow into the FM layer and expresses the net transfer of spins from the FM, the variation of $g_{eff}$ with $t_{Cu}$ in Fig.\ref{Fig4}a validates the fact that spin transmission from Py layer into the adjacent non-magnetic layer has been enhanced by the insertion of Cu layer. Having known the $g_{eff}$, we have calculated the spin current density ($j_s^0$) injected from FM layer into the TI layer by using equation, $j_{s}^{0}=(g_{eff}^{\uparrow\downarrow}\gamma^2 h_{m}^{2}\hbar[M_{eff} \gamma+\sqrt{( M_{eff})^2 \gamma^2+4\omega^2}])/(8\pi\alpha^2[(M_{eff})^2\gamma^2+4\omega^2])$ \cite{geff}. The evaluated $j_s^0$ and the measured $V_{Sym}$ value were used to calculate the inverse Edelstein effect length, $\lambda_{IEE} = j_C/j_s^0$ \cite{Fert2} where we have calculated $j_C$ from the amplitude of the symmetrical Lorentz voltage, $V_{Sym}$ by using the equation, $j_C=V_{Sym}/Rl$ where $l$ is the length of the sample and $R$ is the resistance of the sample measured using four probe method \cite{R}. From Fig.\ref{Fig4}b, one can clearly see that at room temperature the $\lambda_{IEE}$ value has been enhanced for trilayer samples compared to the BSTS/Py bilayer and the change in Cu thickness tunes the $\lambda_{IEE}$. The enhancement of the $\lambda_{IEE}$ for all the trilayer samples suggest a boosting of the net transfer of spin current into the TI layer by minimizing interfacial exchange coupling between the surface state of the TI and the local magnetic moments of the ferromagnetic (FM) material \cite{Exchange1,L2} due to the insertion of the Cu layer in between the BSTS and Py layer. For a complete understanding of the sensitivity of the $\lambda_{IEE}$ value on $t_{Cu}$ for the trilayer samples and to ensure the effect of any possible exchange coupling on the SCC efficiency ($\lambda_{IEE}$) of the device, one needs a low temperature measurement because the exchange interaction occurs between the surface state of TI (TSS) and the magnetic moment of FM and the TSS completely takes over the conduction when we reach low-temperature region.\\
In the proceeding section, we have tried to understand the specific role of TSS in spin-charge conversion. We conducted low temperature measurements on the following samples: BSTS/Py, BSTS/Cu1(1.5 nm)/Py, and BSTS/Cu2(3 nm)/Py where the surface state of TIs dominate conductivity the most. Throughout the experiment, we have used BSTS thin films of thickness 37 nm as the TI material in the heterostructure. We have measured the temperature variation of resistance, $R$ of the BSTS sample which is shown in Fig. \ref{Fig5}a. We have analysed the temperature dependence of conductance to estimate the contribution of surface states using a parallel resistor model following  Singh \textit{et al.} \cite{BSTS2}. We find a $74\%$ contribution of the TI surface state to the overall conductivity and a minimum leakage of spin current into the bulk of the 37 nm BSTS film. We have measured $V_{Sym}$ directly by probing the TI material. From  Fig.\ref{Fig5}b one can see that the temperature variation of measured voltage for all the samples follow the same trend and it is concomitant with the resistance ($R$) vs. $T$ plot shown in Fig.\ref{Fig5}a. As the bulk resistivity increases with lowering temperature and surface conductivity becomes stronger, we are obtaining more voltage due to larger spin accumulation at the TSS. It again establishes the fact that the majority of the spin to charge conversion actually happens at the surface state of TI and not in the bulk of TI and the temperature dependence of voltage is primarily due to the inverse Edelstein effect (IEE) at the topological surface state.
We have studied the temperature variation of FMR linewidth ($\Delta H$) at a resonance frequency of 4 GHz as shown in Fig.\ref{Fig5}c. One can notice that the $\Delta H$ value shows an exponential increase for all the heterostructures once it reaches the temperature range where $R$ saturates due to the maximum dominance of the surface state of TI. The $\Delta H$ enhancement proceeds monotonically with decrease in temperature and reaches the maximum value at the lowest temperature, 8 K for the sample with $Cu$ spacer layer of thickness 3 $nm$. The $g_{eff}$ vs. $T$ shows a similar trend as that of $\Delta H$ as shown in the inset of Fig.\ref{Fig5}b. From the temperature variation data of both the parameters, $g_{eff}$ and $\Delta H$, we can state that lowering the temperature enhances the spin pumping from the FM layer into the adjacent non-magnetic layer. Now the question is whether the pumped spins accumulate directly in the TI surface state  and get convert into a charge current via the Edelstein effect or significant losses due to spin memory loss are also involved in the process of spin to charge conversion at the interfaces.\\
In the final section of our study, we investigated the temperature dependence of  $\lambda_{IEE}$ to understand the role of interface of TI/FM on the spin-charge conversion efficiency of the topological surface state. From the Fig.\ref{Fig5}d we can see that as temperature decreases from room temperature, the value of $\lambda_{IEE}$ initially rises, but then decreases further as the temperature is lowered below a certain point. Spin memory loss at the interface of with the ferromagnetic metal (FM) could be a possible reason behind poor spin to charge conversion at TSS at low temperature. The theoretical model by V. P. Amin \textit{et al.} \cite{S2} predicts that the interface transparency for spin transfer and the spin memory loss at interface depend on the interfacial SOC and interfacial exchange interaction respectively. Our previous study \cite{Pal} with $BiSbTe_{1.5}Se_{1.5}/Ni_{80}Fe_{20}$ and other studies \cite{OS1, OS2} with other TI-based heterostructure suggest that TI/FM bilayers have high interfacial SOC and exchange coupling exists between the TI surface states and the local moments of FM. Spin memory loss that correlates the presence of exchange coupling strength can be significant when we lower the temperature below the point where the topological surface state completely takes over the conductivity. But one can suppress the exchange coupling by a proper separation of the TI and FM layer as can be seen for the sample with Cu thickness of 3 nm. For $BSTS/Cu2/Py$ sample the $\lambda_{IEE}$ value no longer decreases with temperature, indicates the fact that spin memory loss due to presence of exchange coupling got reduced and we achieved $\lambda_{IEE}$ as large as 1.1$\pm$0.03 nm at 8 K and $0.8\pm0.024$ nm at room temperature. The $\lambda_{IEE}$ value for our devices, especially for $BiSbTe_{1.5}Se_{1.5}/Cu$(3 nm)$/Ni_{80}Fe_{20}$ are significantly higher in comparison to the other bilayer samples reported previously \cite{High1,High2,High3,Optical} at lower temperatures as well as at room temperature. Thus to enhance the efficiency of TI-based heterostructures, a suitable choice of the interface layer is crucial. This choice serves to protect the unique spin texture of the topological surface state by reducing the exchange coupling between the TSS and the magnetic layer.\\
In summary, we have observed a noteworthy improvement in the inverse Edelstein effect (IEE) length, which can be controlled by adjusting the thickness of the copper (Cu) layer positioned between the $BiSbTe_{1.5}Se_{1.5}$ and $Ni_{80}Fe_{20}$ layers. By conducting FMR and spin-pumping measurements on $BiSbTe_{1.5}Se_{1.5}/Cu/Ni_{80}Fe_{20}$, we determined a substantial $\lambda_{IEE}$ value of 289 $pm$ for $BiSbTe_{1.5}Se_{1.5}/Ni_{80}Fe_{20}$ structures. Upon introducing a Cu layer at the interface of the topological insulator (TI) and ferromagnetic (FM) layer ($BiSbTe_{1.5}Se_{1.5}/Cu(3nm)/Ni_{80}Fe_{20}$), we discovered that the $\lambda_{IEE}$ can be enhanced further. Specifically, at low temperatures (8K), the insertion of Cu leads to a remarkably increased $\lambda_{IEE}$ value of $(1.1 \pm 0.03) nm$.  Skillful control over the Cu layer will enhance the efficiency of the device by reducing the spin memory loss at the interface that was caused by the presence of exchange interaction of the topological surface state and the local moments of ferromagnetic metal at the interface. The similarity in the temperature dependence of $V_{Sym}$ with that of resistivity implies that the contribution to inverse Edelstein effect at low temperature is dominated by SCI from TSS. The topological protection of surface states increases the potential of the TI/FM bilayer for device application. In essence, our study offers promising ways for designing more efficient spin-charge conversion devices of the topological insulator based heterostructures by controlling the interface effects.

\section*{ACKNOWLEDGEMENTS}
The authors sincerely acknowledge the Science and Engineering Research Board (SERB) (grant no: EMR/2016/007950), the Department of Science and Technology (grant no. DST/ICPS/Quest/2019/22), the DST-INSPIRE, the Council Of Scientific and Industrial Research(CSIR) and the University Grant Commission (UGC) of the Government of India for financial support.


\end{document}